\documentclass[a4paper,11pt]{article}
\pdfoutput=1 

\usepackage{jinstpub} 

\usepackage{lineno}
\usepackage{comment}
\usepackage{xcolor}
\usepackage[normalem]{ulem}
\usepackage{siunitx}
\newcommand{\epsr}[0]{\mbox{$\varepsilon_\mathrm{r}$}}

\newcommand{\CW}[1]{{\color{black}{#1}}}

\newcommand{\SP}[1]{{\color{black}{#1}}}

\newcommand{\partdiff}[2]{\dfrac{\partial{#1}}{\partial{#2}}}

\title{\boldmath Electro-purification studies and first measurement of relative permittivity of TMBi}

\author[a,1]{S. Peters} 
\author[b,2]{B. Gerke,}
\author[a]{V. Hannen,}
\author[a]{C. Huhmann,}
\author[a]{N. Marquardt,}
\author[b]{K. Schäfers,}
\author[c,d]{D. Yvon,}
\author[c,d]{V. Sharyy,}
\author[a]{C. Weinheimer}

\affiliation[a]{Institute for Nuclear Physics, University of Münster, Wilhelm-Klemm-Straße, Münster, Germany}
\affiliation[b]{European Institute for Molecular Imaging, University of Münster, Waldeyerstraße, Münster, Germany}
\affiliation[c]{IRFU, CEA, Université Paris-Saclay, Bâtiment 141, Gif-sur-Yvette, France}
\affiliation[d]{BioMAPs, Service Hospitalier Frédéric Joliot, CEA, CNRS, Inserm, Université Paris-Saclay, place du général Leclerc, Orsay, France}

\emailAdd{peters.simon@wwu.de}

\abstract{ 
A new type of detector for positron-emission tomography (PET) has been proposed recently, using a heavy organo-metallic liquid - TriMethyl Bismuth (TMBi) - as target material. TMBi is a transparent liquid with the high Z element Bismuth contributing 82\% of its mass.  511$\,$keV annihilation photons are converted efficiently into photo-electrons within the detector material producing both Cherenkov light and free charge carriers in the liquid. While the optical component enables a fast timing, a charge readout using a segmented anode can provide an accurate position reconstruction and energy determination. The charge measurement requires a high level of purification, as any electronegative contaminants cause signal degradation. In addition to the purity requirements, the reactive nature of TMBi poses many challenges that need to be met until a fully functioning detector for PET applications can be realized. The paper presents an experimental setup that aims to remove electronegative impurities by electrostatic filtering and to characterise the properties of TMBi, e.g. the relative permittivity, for its application as a detector medium for charge read out.}


\begin{document}
\maketitle
\flushbottom

\section{Introduction}
\label{sec:intro}
\begin{figure}[h]
 \centering
 \includegraphics[width=.95\textwidth]{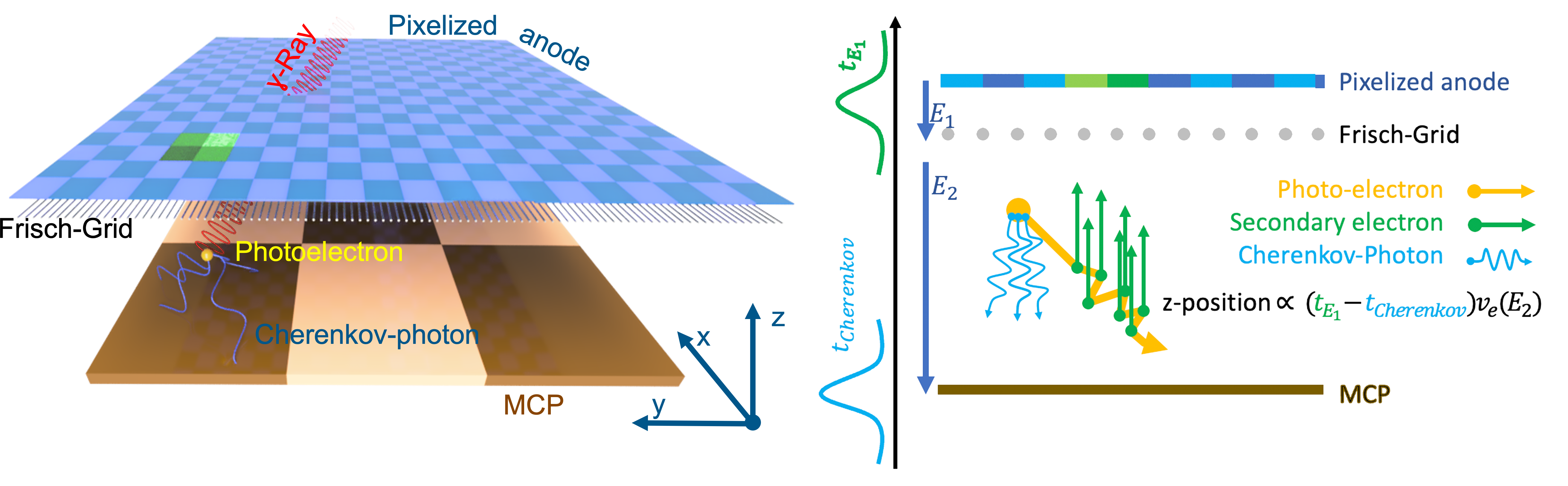}
 \caption{\label{fig:calipso-scheme} The proposed detector combines two readout systems to reconstruct ionization events within three dimensions and with high temporal resolution. Photo-electrons created by 511\,keV quanta in the medium emit Cherenkov radiation that can be detected by fast MCPs, providing a sharp timing signal. In addition the photo-electron further ionizes the detector material, generating secondary electrons which  are drifted by an applied electric field towards a segmented anode, inducing an electric signal that allows a spatial reconstruction of the event in x- and y- direction. Combining the drift time of the secondary electrons with the timing signal of the Cherenkov radiation in enables determination of the z-coordinate (depth of interaction).}
\end{figure}

%
%
The BOLD-PET (Bismuth Organometallic Liquid Detector - Positron Emission Tomography) project aims to develop a detector for PET applications based on the organo-metallic liquid TriMethyl Bismuth (TMBi, chemical composition Bi(CH$_3$)$_3$). 
While the PET technique provides the unique capability to visualize molecular processes in living tissue, the spatial resolution of most clinical PET detectors is quite modest compared to other imaging techniques like CT or MRI. PET utilizes the coincident measurement of two 511\,keV photons emitted back to back in the annihilation of a positron from a $\beta^+$-emitter and an electron from the tissue material. By accumulating many such events a 2D or 3D reconstruction of the activity distribution within the tissue can be achieved. The $\beta^+$-emitter is usually embedded in a tracer molecule administered to the patient. To improve image quality and reduce radiation exposure to the patient it is crucial to detect the annihilation photons with high efficiency and the highest possible energy, temporal and spatial resolution. \\
While most modern clinical PET detectors use inorganic scintillators with good energy resolution but low granularity for that purpose, BOLD-PET pursues a novel detector concept first described in~\cite{Yvon2014}. It will use TMBi, a transparent, dielectric liquid with a density of 2.3\,g/cm$^3$, and Bismuth as a high Z component for efficient photo detection. As seen in figure~\ref{fig:calipso-scheme} the detector medium is placed between two readout systems, a segmented anode for charge readout allowing spatial and energy reconstruction, and an ultra-fast MCP array with sub nanosecond temporal resolution for photon counting.
Bismuth, being the heaviest non radioactive element, accounts for 82\% of the weight in TMBi \cite{Yvon2014,ramos}. It converts 511\,keV quanta efficiently into relativistic photo-electrons. Due to the high refractive index of TMBi ($n_{\rm TMBi}=1.63\;@\;\lambda = 425\,$nm) photo-electrons generated in TMBi propagate faster than the phase velocity of light within the dielectric liquid, thereby emitting Cherenkov radiation~\cite{ramos}. Using ultra-fast MCPs, read out by SAMPIC digitization modules~\cite{Canot}, the radiation is detected with a temporal resolution $<150\,$ps allowing for TOF-PET applications \cite{calipso_sym}. Until it is fully thermalized, the primary electron will create further electron ion pairs in the detector material. An electric field applied between the MCP plane (cathode) and the pixelized anode causes electrons and ions to drift in opposite directions towards their respective electrodes inducing image charges. However, only the signals of the fast moving electrons will be detected using charge sensitive preamplifiers. To eliminate the dependence of the signal height on the start position of the electrons, a Frisch grid on an intermediate electric potential is placed close to the anode. In that way only the drift length between the Frisch grid and the anode, which is the same for all electrons created between cathode and grid, has an impact on the signal height.\\
\CW{Neglecting electron losses while drifting by attachment at electronegative impurities,} the charge measured at the anode then only depends on the amount of energy deposited in the liquid and on the field dependent free ion yield $G_{fi}$ of TMBi which determines the amount of electrons generated per deposited energy. In addition to measuring the incoming photon's energy the pixelized anode enables determination of the x- and y- positions of the photoionization event within the detector. By also measuring the time delay between the charge signal obtained at the anode and the fast timing signal obtained from the MCP one can calculate the depth of interaction (z-position) within the detector. \\
Electronegative impurities in the detector material can trap free electrons, forming slow moving ions that are not detected by the readout electronics and thus reduce the energy \CW{signal}. It is therefore crucial to reduce the presence of impurities in the liquid to an absolute minimum.
\CW{The high density of a liquid medium in a drift detector imposes significantly higher purity requirements than those for a drift detector with a gaseous medium. To estimate the necessary detector purity with respect to electronegative impurities, it is certainly reasonable to require that the mean free drift distance $\lambda_\mathrm{d}(\mathrm{TMBi})$ is comparable to the attenuation length $\lambda_\mathrm{a}(\mathrm{TMBi,511\,\mathrm{keV}}) \approx 3\,\mathrm{cm}$ of 511\,keV annihilation photons in TMBi \cite{xcom}. Unfortunately, this constraint still does not allow us to specify the purity requirements for the TMBi because we do not know all the properties of TMBi yet. Therefore, to estimate the purity requirement for TMBi, we use the corresponding numbers of TMS (Tetramethyl-Silane), also an organometallic detector medium that has been studied in detail in many detector applications.}

\CW{Drifting electrons may get lost by attachment $\mathrm{e}^- + \mathrm{S} \rightarrow \mathrm{S}^-$ on impurities S with a impurity concentration $c_\mathrm{S}$ described by an attachment coefficient $k_\mathrm{S}$:
\begin{equation} \label{eq:e_attachment}
  \frac{\partial c_\mathrm{e}}{\partial t} = - k_\mathrm{S} \cdot  c_\mathrm{S} \cdot c_\mathrm{e}
\end{equation}
leading to an exponential decrease of the electron concentration $c_\mathrm{e}$ characterized by the electron drift lifetime $\tau = (k_\mathrm{S} \cdot  c_\mathrm{S})^{-1}$. This electron drift lifetime $\tau$ times the electron drift velocity $v_\mathrm{d}$, given by electron mobility $\mu_\mathrm{e}$ and electric field strength $E$, specifies the mean free drift distance $\lambda_\mathrm{d}$. The latter we requested to be larger than the attenuation length of annihilation quanta in TMBi $\lambda_\mathrm{a}$:
\begin{eqnarray}
      \lambda_\mathrm{a}(\mathrm{TMBi,511\,\mathrm{keV}}) & < & \lambda_\mathrm{d} = v_\mathrm{d} \cdot \tau = \frac{\mu_\mathrm{e} \cdot E}{k_\mathrm{S} \cdot  c_\mathrm{S}}\nonumber \\
      \Leftrightarrow c_\mathrm{S} & < & \frac{\mu_\mathrm{e} \cdot E}{k_\mathrm{S} \cdot  \lambda_\mathrm{a}(\mathrm{TMBi,511\,\mathrm{keV}})} \label{eqn:impurity_constraint}
\end{eqnarray}
Assuming oxygen as the main impurity S and putting into Eqn. (\ref{eqn:impurity_constraint}) the corresponding numbers for TMS,  $k_\mathrm{O}(\mathrm{TMS})=6 \cdot 10^{11}\,\mathrm{l/(mol\,s)}$ and 
$\mu_\mathrm{e}(\mathrm{TMS}) = 98\,\mathrm{cm^2/(V\,s)}$
from reference \cite{Holroyd}, we obtain for a field strength $E=3.33\,\mathrm{kV/cm}$:
\begin{equation}
    c_\mathrm{S} < 
    1.8 \cdot 10^{-7}\,\mathrm{mol/l} \label{eq:impurity_limit}
\end{equation}
With a molar density of TMBi of 
$9.1\,\mathrm{mol/l}$ this corresponds to our estimated request for the maximum relative oxygen concentration of 20\,ppb.
 }

While commonly used methods like filtering using molecular sieves or distillation of the liquid to remove impurities have also been applied by the authors, in this paper we will focus on so-called electrostatic filtering. The method has been proposed by Kai Martens (Kavli-IPMU, U. Tokyo) and aims to make use of the electron affinity of the contaminants which are charged up by providing surplus electrons to the liquid and then removed by an electric drift field~\cite{Martens}. 

The paper is structured as follows: in section~\ref{sec: chapter_purification_Plus_detector} we discuss the experimental setup used to perform the measurements. Section~\ref{sec:purificationstudies} describes the TMBi purification tests performed with emphasis on the electrostatic filtering and section~\ref{sec:relativePermittivity} provides the results of the relative permittivity measurement.

\section{Experimental setup}
\label{sec: chapter_purification_Plus_detector}
\begin{figure}[h]
 \centering
 \includegraphics[width=.6\textwidth]{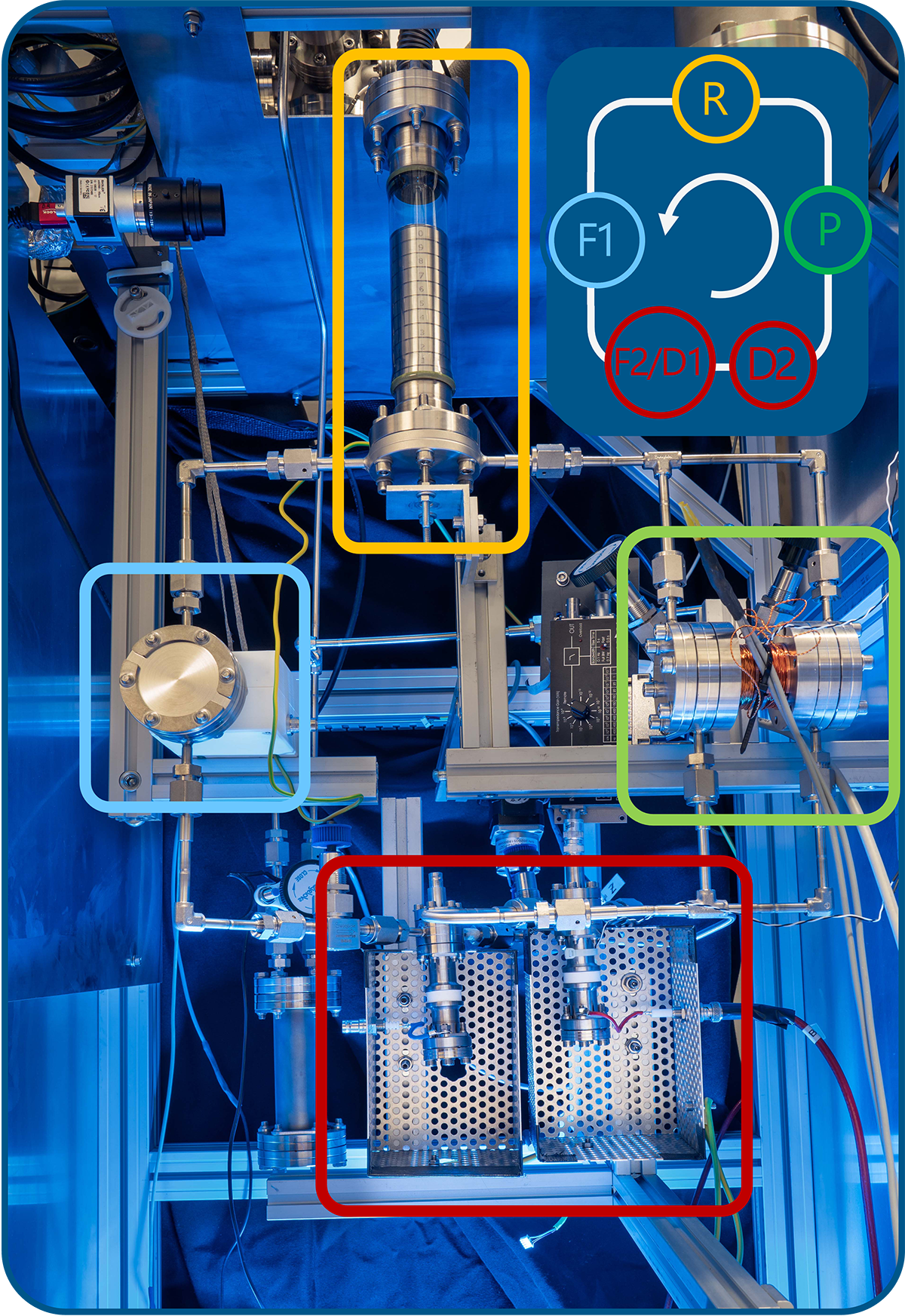}
 \caption{Overview of the main components of the TMBi test setup: (R) Reservoir, (F1) electrostatic filter, (F2/D1) combined filter / detector cell, (D2) detector cell and (P) magnetically driven piston pump.}
 \label{fig: electro pur overview} 
\end{figure}

The reactive nature of TMBi, which spontaneously decomposes at temperatures above $105^{\circ}$\,C, poses strict limitations on the materials used in the test setup and the environmental conditions applied \cite{Gerke:2022otq}. 
The setup presented in this paper therefore makes use of ultra-high-vacuum grade components built exclusively from stainless steel, ceramics (Al$_2$O$_3$), glass and copper. This minimizes the risk of chemical reactions between the TMBi and the apparatus as well as the introduction of impurities. Additionally, it hermetically seals the liquid from the environment. 
Figure~\ref{fig: electro pur overview} shows a photograph of the test setup.
Its main part consists of a loop of stainless steel pipes and full metal valves connecting a TMBi reservoir (R), an electrostatic filter (F1), two detector cells (D1) and (D2) and a small piston pump (P) to move the TMBi through the loop. 
The detector cell (D1) is alternatively also used as an electrostatic filter and then referred to as (F2). Not shown in the photo is the required additional infrastructure like temperature controlled TMBi containers, a turbo-molecular pump, a dry pre pump, cold traps, pressure gauges, etc. Before the reservoir is filled with TMBi, all components are evacuated and purified by baking at a temperature of 200$^{\circ}$\,C, with the exception of metal to glas/ceramic transitions, which were baked at lower temperatures within the manufacturers specifications. \\ 
Applying electric fields to the liquid \SP{at the saturated vapor pressure of TMBi (40\,mbar at 20$^{\circ}$C)}, where the boiling point of TMBi is close to  room temperature, increases the risk of forming gas bubbles between the electrodes which in turn can lead to electrical discharges and sudden decomposition of the TMBi.
To reduce this risk, highly purified argon gas is used to pressurize the liquid at $\approx 1\,$bar after initial filling. 
This precaution has been shown to reliably suppress explosive behavior in separate test measurements described in~\cite{Gerke:2022otq}.\\
The following sections describe the above mentioned main components of the TMBi loop. 

\subsection{Reservoir}
\label{sec:reservoir}

The Reservoir is constructed from a CF-40 glass cylinder of 200\,mm length. On top of the cylinder a standard CF-40 flange is used to couple the test setup shown in figure~\ref{fig: electro pur overview} via a bellow to the rest of the TMBi handling infrastructure. The bellow reduces mechanical stress exerted on the glass-metal transition. Inside the vessel a full-metal cylinder displaces \SP{approximatly 100\,ml of the total reservoir volume of 170\,ml, reducing the amount of TMBi required for operation}. The glass enclosure allows to visually inspect the liquid level, the \SP{the formation of a meniscus during pump operation, which is used to estimate the pump velocity, } and the optical quality of the TMBi which proved very useful during the measurements. An engraved scale allows to measure the amount of TMBi that is extracted. The reservoir is located upstream of the electrostatic filter.

\subsection{Electrostatic filter}

\label{sec: description_ES-Filter}
\begin{figure}[h]
 \centering 
 \includegraphics[width=.95\textwidth]{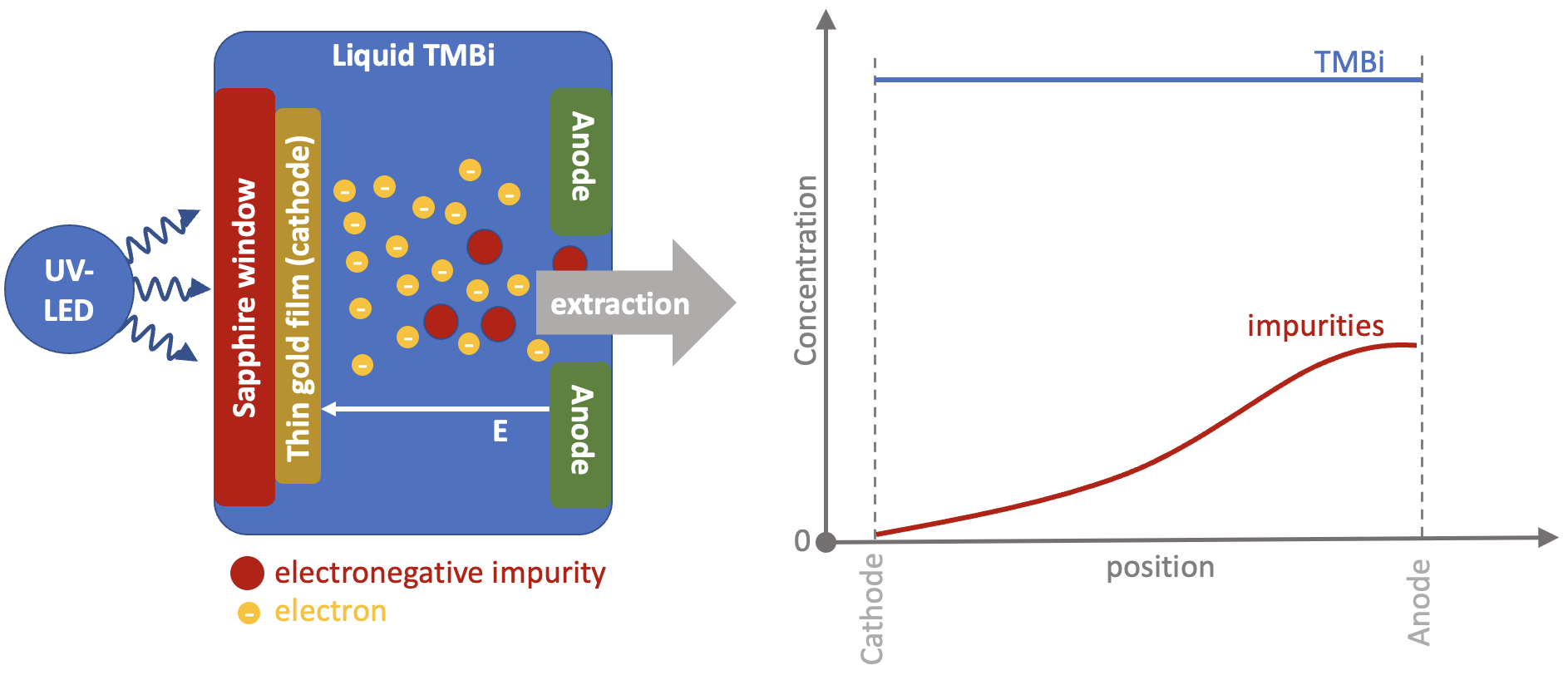}
 \caption{\label{fig: FilterFunction} Our idea how an electrostatic filter should affect the concentrations of electronegative impurities. Injected electrons, e.g. by photoelectric effect, charge up electronegative impurities that are guided by an electrostatic field towards the anode. Removing material in the region of the highest impurity concentration (in front of the anode) will result in an overall impurity reduction. }
\end{figure}

\begin{figure}[h]
 \centering 
 \includegraphics[width=.95\textwidth]{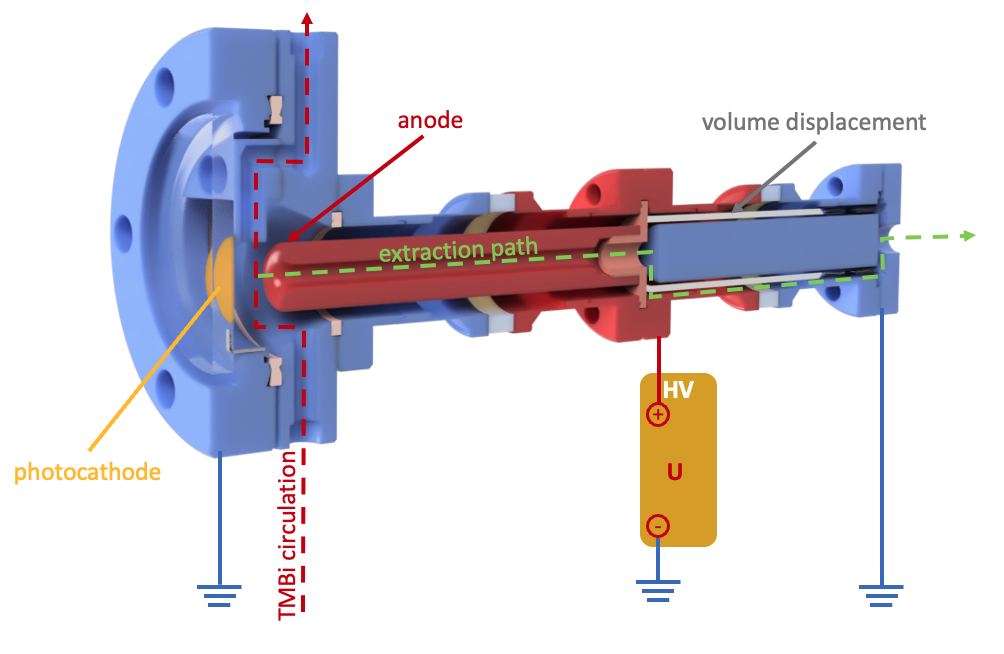}
 \caption{\label{fig: FilterScheme} Schematic view of the electrostatic filter F1. A UV-LED array illuminates a photo cathode on the inside of a UV transparent window. The light creates free electrons from the cathode material which are drifted through the TMBi by a positive high voltage applied to the anode. The gray volumes represent ceramic insulators. An opening in the anode allows to extract small amounts of the liquid \SP{into a separated vacuum container}.}
\end{figure}

The idea of purifying TMBi in an electrostatic filter (ESF) works by providing free electrons to the liquid in a plate-capacitor like setup, as seen in figure \ref{fig: FilterFunction}. 
The electrons will predominantly be captured by \SP{electronegative impurities} which can then be manipulated by applying an electric field between the electrodes. The negatively charged ions will drift towards the positively charged anode, where they are neutralized and diffuse back into the volume. A strong electric field and high enough flux of free electrons injected at the cathode, e.g. generated by a photo cathode, will force impurities to accumulate faster at the anode than they can diffuse back into the volume, resulting in a concentration gradient of impurities with the highest concentration at the anode. Removing liquid in this region should then reduce the concentration of contaminants in the remaining detector material.
\CW{To just give a rough estimate of the required free electron current density $j$ for an electrostatic filter let us assume at the start of the purification a factor 10 higher impurity concentration of $c_S=1.8\cdot 10^{-6}\,\mathrm{mol/l}$ or in impurity number concentration $n_S=1.1 \cdot 10^{15}\,\mathrm{/cm^3}$ than our request in Eqn. \ref{eq:impurity_limit}. For an electrostatic filter of length $l=\lambda_\mathrm{a}(\mathrm{TMBi,511\,\mathrm{keV}}) \approx 3\,\mathrm{cm}$ this corresponds to an impurity column density of $n_S \cdot l=3.3\cdot 10^{15}\,\mathrm{/cm^2}$. A free electron current density of $1\,\mu\mathrm{A/cm^2}=6.3 \cdot 10^{12}\,\mathrm{e/s\,cm^2}$ will then require about $9\,\mathrm{min}$ to attach an electon on every impurity atom/molecule, thus drifting all impurities within the electrostatic filter in that time once to the anode in order to take them out.}\\
Figure \ref{fig: FilterScheme} shows a schematic view of the electrostatic filter developed for the purification test setup. 
It contains a photo cathode consisting of a thin gold layer ($\approx 20\,$nm) that is evaporated onto the vacuum side of a standard UV-transparent CF-40 silica glass flange. For increased mechanical stability, a $1\,$nm thick titanium film is used as an adhesive layer between glass and gold film. A gold plated copper ring secured with a horse-shoe shaped stainless steel frame inside the CF-40 window is used to ensure electrical contact between the flange (on ground potential) and the gold cathode. The coated window is mounted on a custom made CF-40 flange that connects the ESF to the TMBi loop via two $1/4$" VCR connectors. On the back side the flange connects to a pair of CF-16 insulators, in the middle of which a cylindrically shaped anode is mounted. The latter is connected to a positive high voltage, supplied by a programmable power supply iseg NHQ 224M (all parts on positive potential are shaded red in figure~\ref{fig: FilterScheme}). 
TMBi that is circulated in the setup is guided through the volume between photo cathode and anode where the highest flux of photo electrons is generated. The anode is an $80\,$mm long cylinder with a parabolically shaped extraction channel placed at a distance of $10\,$mm from the photo cathode. 
This configuration allows to extract material directly in front of the anode, where the highest concentration of electronegative impurities is expected when the photo cathode is illuminated by an array of 24 SMD (Surface-mounted device) mounted UV-LEDs (255~nm wavelength) through the window.  
A motorized valve located downstream of the anode can be used to extract small amounts of liquid at the tip of the anode \SP{into a separate vacuum container}. A stainless steel rod isolated by a glass cylinder and positioned between anode and valve is used to reduce unnecessary volume. The amount of liquid extracted can be monitored externally on the gauge engraved in the reservoir.

\subsection{Detector}

\label{sec: description_detector}
\begin{figure}[h]
 \centering
 \includegraphics[width=.95\textwidth]{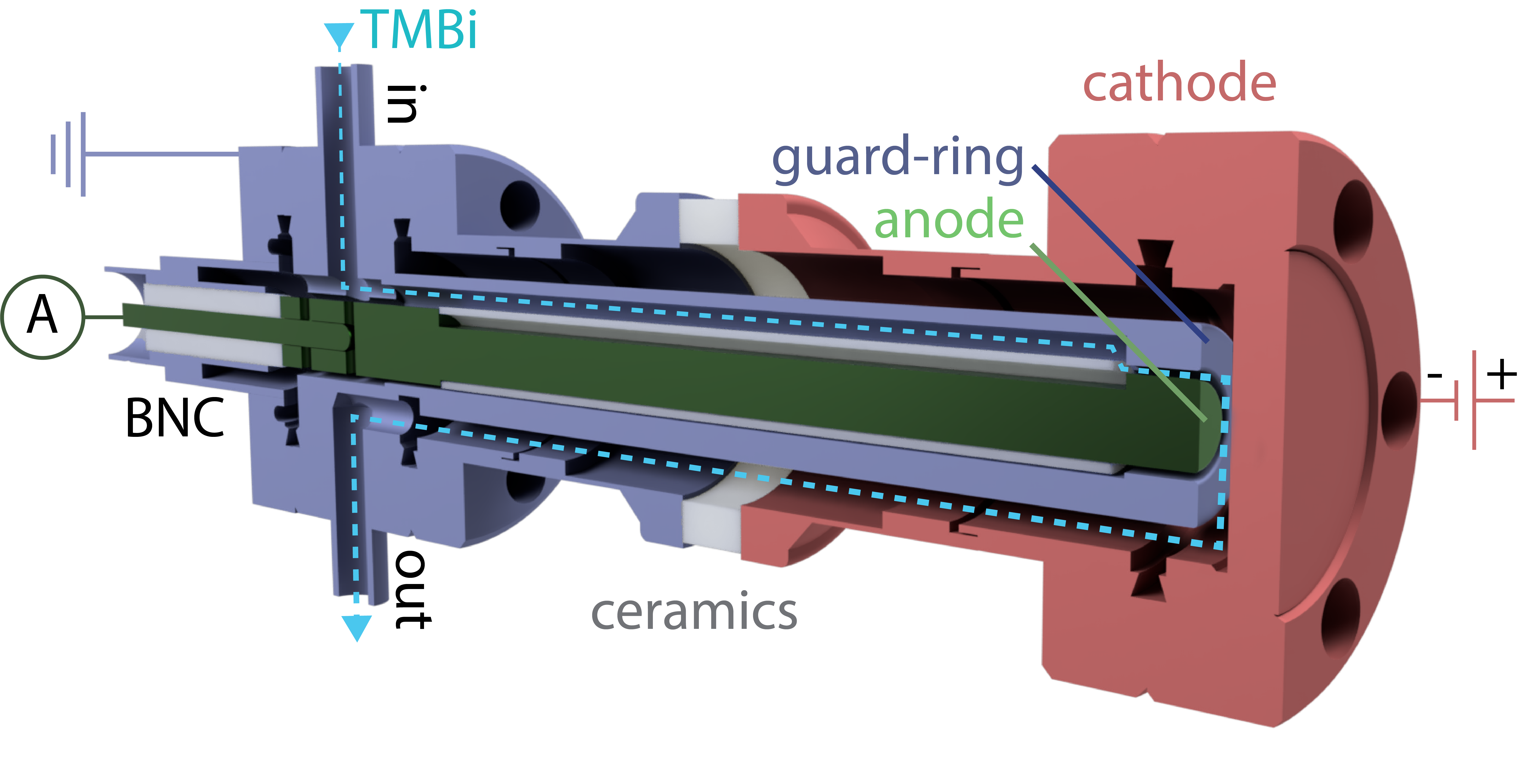}
 \caption{Schematic view of the detector cell \SP{D1(F2) and D2} for the TMBi test measurements. Liquid enters the detector from the top and is guided towards the interface between anode and cathode. Blue shaded parts in the drawing are grounded while a positive high voltage can be applied to the red parts. The anode (green) is connected to the measurement electronics via a BNC feedthrough. }
 \label{fig: detector scheme}
\end{figure}

The purification process is monitored using a detector (see figure~\ref{fig: detector scheme}) functioning as a conductivity cell located downstream of the ESF. 
TMBi can be pumped through the active volume of the detector, enabling online measurements of the conductivity of the liquid which relates to the purity. The design is based on a simple plate capacitor geometry with an additional guard ring to reduce electric field inhomogeneities between anode and cathode.
The anode consists of a stainless steel rod isolated by a ceramic tube inside a stainless steel cylinder of $78\,$mm length. The tip of the anode forms a level surface with the face of the tube that acts as the guard ring. The guard ring is set to ground potential, whereas the anode is virtually grounded via a current amplifier (DDPCA300, FEMTO, Germany). Its output is digitized using a Voltage Input Module (NI 9205, NI, USA). A CF16-BNC\footnote{CF16-BNC-GS-SE-CE-SS} feed through is used to electrically connect the anode to the measurement electronics.
A CF-16 insulator piece\footnote{CF16-ISO6M-CE-KOV190} forms the outer shell of the detector, serving as container and galvanic isolation of the cathode from the rest of the system. The cathode is simply realized by a CF-16 blind flange mounted at the right end of the insulator and connected to a positive voltage supply. To reduce the read out noise we typically use a 9\,V battery to provide the cathode voltage. The flange can easily be replaced or modified to adapt the detector to specific needs e.g. for testing a Frisch-grid between anode and cathode or for implementing a photo cathode that allows to study the electron mobility or electron lifetime in the liquid. 

\subsection{Miniaturized magnetically driven piston pump}
\label{sec: description_pump}
\begin{figure}[h]
 \centering
 \includegraphics[width=\textwidth]{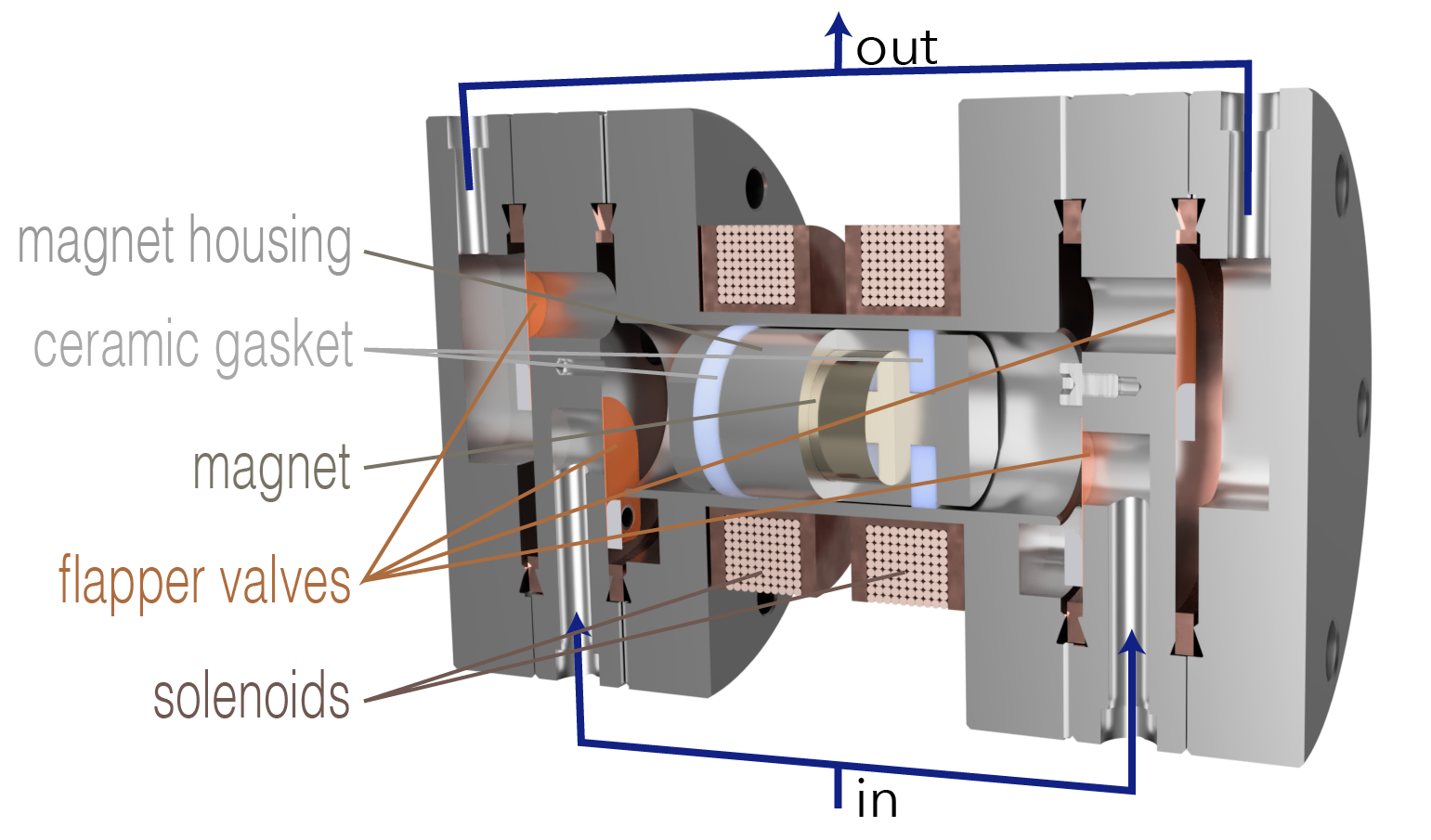}
 \caption{Miniaturized magnetically driven piston pump.}
 \label{fig:pumpe}
\end{figure}

Continuously circulating TMBi through the system allows to directly monitor the purification process. The protective argon atmosphere, that pressurizes the liquid for safe operation when a voltage is applied to the ESF prohibits thermal pumping of the liquid. Instead, a small volume liquid pump is used to operate the system. The reactive nature of TMBi and the high requirements on purity exclude the use of commercially available small volume liquid pumps. Therefore, a miniaturized magnetically driven piston pump, designed to meet our high vacuum and purity requirements, was developed and manufactured in the institute workshop. The design is influenced by high-purity magnetically driven pumps developed in our group for the XENONnT experiment~\cite{Brown} with a focus on using UHV- and TMBi-compatible materials and miniaturization of the system to reduce the total amount of TMBi stored in the setup. 
Figure~\ref{fig:pumpe} shows a cross section of the pump.
The piston consists of a hermetically sealed stainless steel container housing a 15~mm diameter cylindrical neodymium (NdFeB) magnet with an internal magnetic flux of $1.32\,$T. It is actuated by two external solenoids coils made from 0.5mm thick enamelled wire, with approx 280 windings each. The coils are wound onto copper bobbins that also serve to transfer heat away from the liquid. 
As materials like PE or Teflon can not be used when working with TMBi, two ceramic disks\footnote{FRIALIT®-DEGUSSIT® High Performance Ceramics} with a diameter of 20\,mm are attached on both ends of the piston serving as gaskets. The inner diameter of the 78\,mm long tube is honed to fit the dimensions of the ceramic disks and ensure that the piston can move freely while giving a good enough seal. Both tube endings are capped with custom CF-40 flanges that enable the liquid to enter and leave the pump through flapper valves constructed from sub-millimeter spring-steel. The correct positioning of the valves on the in- and outlet flanges ensures an unidirectional flow. Not including tubes, the Pump has an overall volume of approximately 33~ml and can move up to 5\,ml per stroke. Smooth actuation of the piston is achieved by driving one coil at a constant current, repelling the piston, while a sinusodial current on the other coil alternates between attracting and repelling the piston. 

\section{Purification Studies} 
\label{sec:purificationstudies}
After pre-cleaning of the TMBi and filling into our setup described in section 2 we present in this section first electropurification measurements. In the first campaign we used the dedicated electrostatic filter F1 to remove electronegative impurities and measure the purification effect in the TMBi with detector D2. In the second campaign we used the second detector as an electrostatic filter F2 and again measure the purification effect in the TMBi with the detector D2. 

\SP{
\subsection{Conductivity due to impurities} \label{sec:conductivity_impurities}
Previous measurements using TMBi showed that the liquid has good insulating properties, electric currents in the fA range due to ionizing radiation could be measured 
\cite{Farradeche}. Also our previous measurements using TMS in similar detectors showed no significant effects of self-conductivity, validating our measurement methods \cite{master_simon_peters}.
We therefore assume that the large conductivity observed here with our TMBi is due to some unknown impurities $S$ (likely, not necessarily the ones which are responsible for the electron attachment in Eqn. (\ref{eq:e_attachment})). 
Similar self-conductivity effects were previously observed in organic liquids like cyclohexan: In reference \cite{Staudhammer} the measured current at a given voltage and thus  the conductivity $\sigma$ of an non-irradiated detector was found to linearly depend on the impurity concentration of water. To lead to a non-zero conductivity by the impurities $S$, they have to give rise to the creation of ions. Several processes are possible depending on the kind of impurity $S$ and the kind of liquid $L$, cyclohexan in the case reported in \cite{Staudhammer}, TMBi in our case. 
These impurities seem to dissociate into positively $S_1^+$ and negatively charged ionic $S_2^-$ with a ionisation rate $k_i$.  The ionic fragments could recombine again with a rate coefficient $k_r$. Also the liquid atoms/molecules $L$ itself may get involved \cite{liquid-state-properties}
\begin{eqnarray}
    S & \overset{k_i}{\underset{k_r}{\rightleftarrows}}& S_1^+ + S_2^-
    \label{eq:1nach2}
    \qquad \text{or} \\
    S + S & \overset{k_i}{\underset{k_r}{\rightleftarrows}}&  S_1^+ + S_2^- \label{eq:2nach2}
    \qquad \text{or} \\
    S + L & \overset{k_i}{\underset{k_r}{\rightleftarrows}}& S^\pm + L^\mp 
    \label{eq:1uLnach1uL}
\end{eqnarray}
The type of reaction defines the dependence of the ion concentration $c_{S^{\pm}}$ on the impurity concentration $c_S$.
Eqn. \ref{eq:1nach2}
leads to a change of the impurity concentration
\begin{equation}
    \partdiff{c_S}{t} = -k_i \cdot c_S + k_r \cdot 
    \underbrace{c_{S_1^+} \cdot c_{S_2^-}}_{c_{S_1^+}=c_{S_2^-}:=c_{S^{\pm}}} = -k_i \cdot c_S+ k_r \cdot c_{S^{\pm}}^2 \label{eq:1nach2rate}
\end{equation}
In the equillibrium case $\partial c_S /\partial t=0$, Eqn. \ref{eq:1nach2rate} leads a ion concentration $c_{S^{\pm}}$, which depends on the square root of the impurity concentration $c_S$:
\begin{equation}
    c_{S^{\pm}} = \sqrt{\frac{k_i \cdot c_S}{k_r}}
\end{equation}
In equilibrium reaction \ref{eq:1uLnach1uL} yields also a square root dependence of the 
ion concentration $c_{S^{\pm}}$ on the impurity concentration $c_S$ because of the high and thus constant liquid concentration $c_L=a$:
\begin{equation}
    0=\partdiff{c_S}{t} = -k_i \cdot c_S \cdot a + k_r \cdot 
    c_{S_1^+} \cdot c_{S_2^-} 
    \quad \text{with } c_{S_1^+}=c_{S_2^-}:=c_{S^{\pm}}:\quad  c_{S^{\pm}} = \sqrt{\frac{k_i \cdot a \cdot c_S}{k_r}} 
\end{equation}
Differently, in equilibrium reaction \ref{eq:2nach2} yields a linear dependence of the ion concentration $c_{S^{\pm}}$ on the impurity concentration $c_S$: 
\begin{equation}
    0=\frac{1}{2} \cdot \partdiff{c_S}{t} = -k_i \cdot c_S^2 + k_r \cdot 
    c_{S_1^+} \cdot c_{S_2^-} 
    \quad \text{with } c_{S_1^+}=c_{S_2^-}:=c_{S^{\pm}}:\quad  c_{S^{\pm}} = \sqrt{\frac{k_i}{k_r}} \cdot c_S 
\end{equation}
This behaviour was reported for water impurities in cyclohexan \cite{Staudhammer}, which might well be connected to the fact, that the well-known protolysis reaction $2 \text{H}_2\text{O} \rightleftarrows \text{H}_3\text{O}^+ + \text{HO}^-$ belongs to the class of reactions described by Eqn. \ref{eq:2nach2}.
Finally we would like to state that 
the electric conductivity $\sigma$ 
due to the impurities is given by the elementary charge $e$, the ion concentrations $c_{S^{\pm}}$ and the ion mobilities $\mu_{\pm}$ 
\begin{equation}
    \sigma = e \cdot c_{S^{\pm}} \cdot (\mu_+ + \mu_-) \propto \left\{  
    \begin{array}{cl}
         \sqrt{c_S} \quad & (\text{reactions~\ref{eq:1nach2},\ref{eq:1uLnach1uL}}) \\
      c_S  \quad & (\text{reaction~\ref{eq:2nach2}}) \\
    \end{array} 
\right. 
\label{eq:conductivity}
\end{equation}
The TMBi used in the experiments presented has been supplied by Chemtura (now Lanxess) with a proclaimed purity of >99.9\%. It was shipped in a metal container with a Teflon seal. Upon inspection of the liquid inside an argon filled glovebox it was observed that the liquid was not fully transparent as expected but appeared foggy. This might indicate a contamination of the TMBi during transport, possibly due to an insufficient seal on the container of the supplier. Either atmospheric gases entered the container or the Teflon-seal itself reacted with the TMBi (Teflon is known to decompose in contact with some metals or hydro carbons~\cite{Teflon,schmidt}). Another possible impurity could be chlorine which is used in the production process of our supplier \cite{patent}. 
}

\subsection{Precleaning of the TMBi}
A combination of degassing, cryo-distillation and drying using 4A molecular sieves was conducted prior to filling the liquid into the purification system, to reduce the introduction of impurities into the apparatus. After transferring the TMBi to the reservoir the system was pressurized with a clean argon atmosphere of $\approx 1$~bar.
\subsection{Method to characterize the electrostatic filters} \label{sec:methods}
The performance of the electrostatic filters can be characterized by measuring the current in the detector D2 at a fixed cathode voltage. This current is a measure of the self conductivity of the TMBi, \CW{which according to section \ref{sec:conductivity_impurities} should be directly correlated to the impurity concentration.\\}
Under vacuum the dark current of the electrostatic filter F1 at a filter voltage of 2\,kV was below the measurement threshold of our power supply of ${\cal O}(100)$\,pA. With full UV irradiation we obtained filter currents of ${\cal O}(1)$\,nA. 
Filling the electrostatic filter with TMBi yielded at the same voltage already a sizeable filter current of ${\cal O}(1)$\,$\mu$A even without UV irradiation of the photo cathode.
Switching on the UV irradiation did not result in a further visible increase of the filter current. \\
Apparently, \SP{due to dissosiation of impurities}, the concentration of ions in the liquid was already high before the additional introduction of charge carriers by the photo-electrons. 
As we also observed problems with the stability of the semi-transparent gold layer (that acts as the cathode) when in contact with a flow of TMBi, we decided to replace the gold coated UV window by a CF-40 blind flange as the new cathode and start a series of measurements to investigate the \SP{self conductivity of} the  \SP{TMBi in an adjacent detector D2} when voltage is applied to F1 \SP{and no additional electrons are injected via a photoelectrode}. \\
To ensure equal starting conditions the TMBi is circulated through the setup with the filter voltage set to zero prior to each measurement. Since no dedicated flow meter is installed in the loop the flows can only be monitored visually through the glass cylinder located at the reservoir. \SP{The outlet of the pump cycle is located on top of the displacement cylinder of the reservoir, resulting in the formation of a small meniscus during pump operation, allowing to estimate the pump velocity.} This rudimentary way of adjusting the flows results in a large variance in pumping velocities, as a consequence the timing constants presented in the following measurements might vary in between runs.

\subsection{First electropurification measurement}
\begin{figure}[b]
 \centering 
 \includegraphics[width=.95\textwidth]{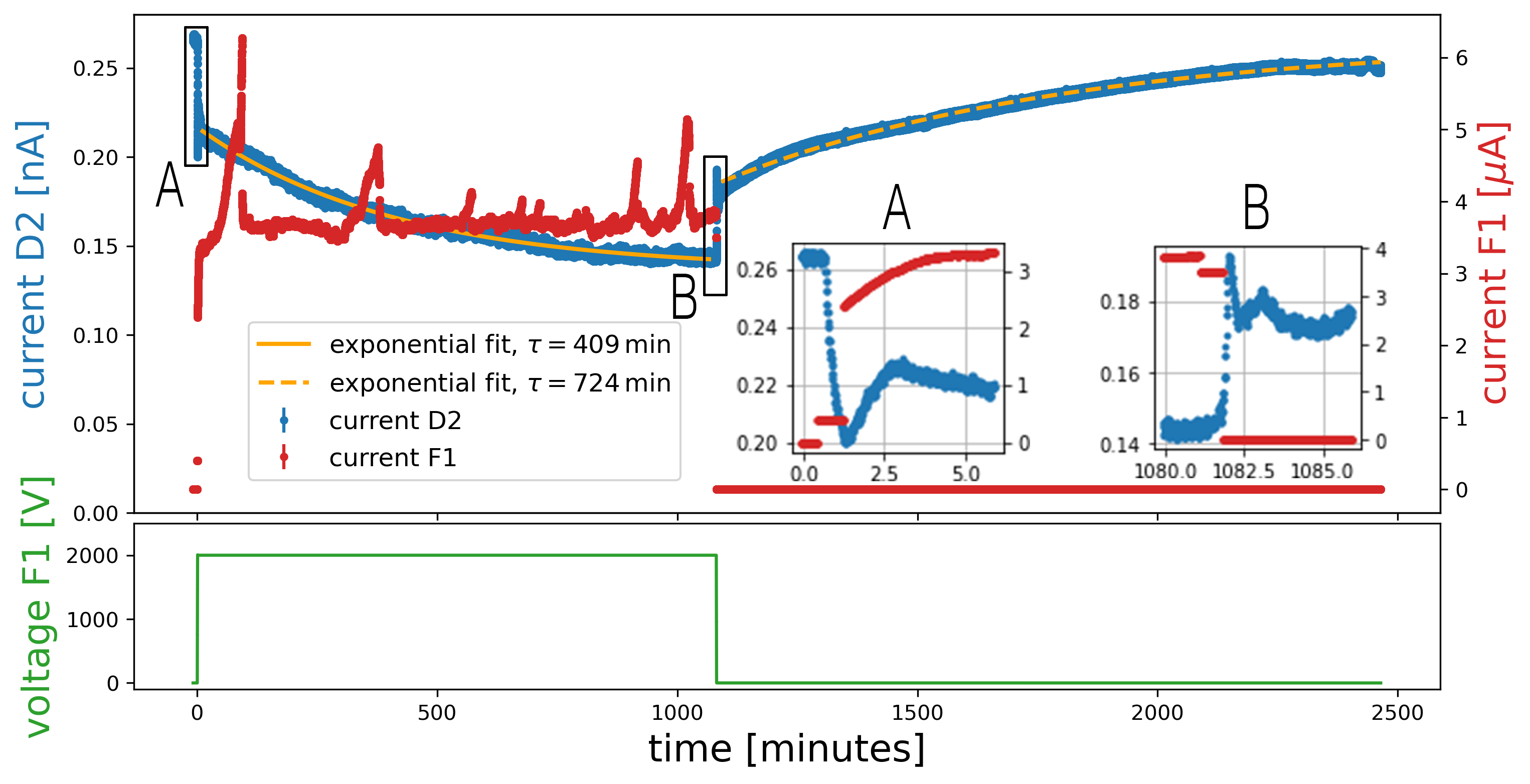}
 \caption{\label{fig: FirstFilter} 
 Test measurement of purification with the electrostatic filter F1 checked with the detector D2.
 Upper panel: current measured at detector D2 (blue) and electrostatic filter F1 (red). Lower panel: voltage applied to F1 (green).}
\end{figure}
Figure \ref{fig: FirstFilter} shows a typical measurement with the detector D2 and electrostatic filter F1. TMBi is circulated for approximately 2500\,minutes through the system and the currents in F1 and the detector D2 are recorded in dependence of the electrostatic filter voltage and time. 
At $t=0$ the anode voltage of the filter F1 is increased from 0\,V to 2000\,V with a ramping speed of 50\,V/s. Approximately 15\,s after the voltage ramp started the current measured at detector D2 begins to drop sharply from 0.26\,nA to 0.21\,nA (see figure \ref{fig: FirstFilter}, insert A). Within the next 1100\,min it then slowly decreases exponentially with a time constant of about 600\,min  to a minimum of 0.14\,nA, while the current measured at F1 exhibits 
a slow, mainly linear increase with some irregular current spikes  with amplitudes of 1-3\,$\mu$A . After approximately 1100\,min the filter voltage at F1 is switched off, while the pump continues to cycle the TMBi and the measurement voltage applied to D\SP{2} stays constant. Switching off F1 results in an inverse behaviour of the current measured at D\SP{2}. After an initial sharp rise (see figure \ref{fig: FirstFilter}, insert B), the current  increases with a similar time constant as it dropped before. A maximum current of $0.25\,$nA is reached 20\,h after switching F1 off. \\
We interpret the fast drop (rise) of the detector current after switching on (off) the filter voltage to originate from a temporary trapping of impurities in the electrostatic filter as indicated in figure \ref{fig: FilterFunction}. The time delay of its occurrence of about 15\,s matches the ratio of the volume between F1 (25\,ml) and D\SP{2} of 10\,ml and the pumping speed of about 1\,ml/s. Similarly the fall (rise) time of the drop (rise) of about 15\,s corresponds well to the volume of F1 of 25\,ml.\\
The much longer time scale of the slow decrease (rise) of about ${\cal O}(500)$\,min presumably has a different origin than the temporary trapping of ions in the filter. Here our hypothesis is that charged impurities are retained electrostatically on the surfaces of the isolators present in the filter. Its long time constant then reflects the attachment rate whereas its exponential behaviour corresponds to a finite trapping capability of the ceramic surfaces.

In summary we observed a significant reduction of the detector current by nearly a factor of 2 when the electrostatic filter F1 is switched on. Since, in this measurement, we did not actively remove TMBi from the filter volume, the impurity trapping was only temporarily. The observed two time constants indicate two trapping mechanisms.  

\subsection{Test of TMBi removal from the electrostatic filter}
\begin{figure}[t]
 \centering 
 \includegraphics[width=.95\textwidth]{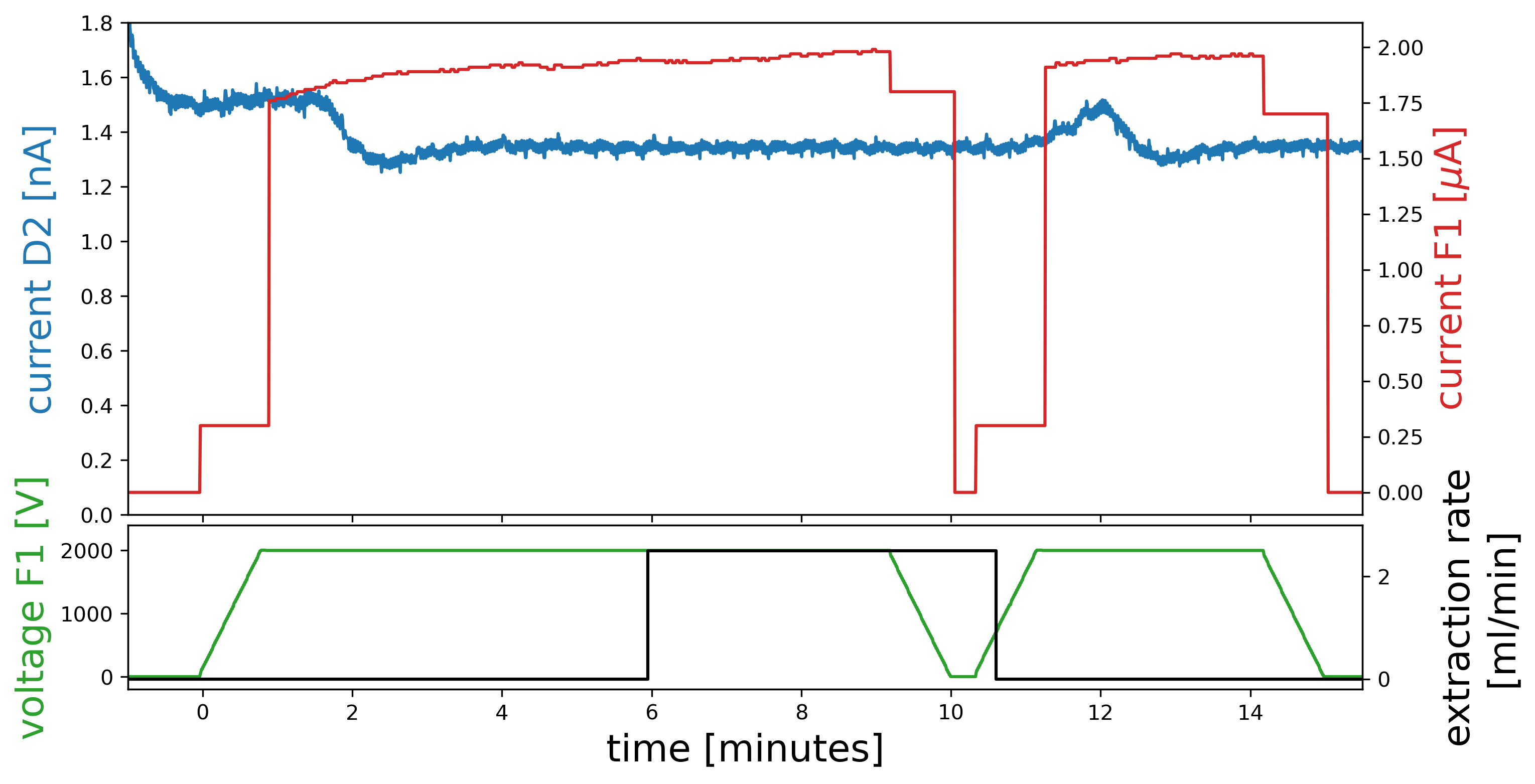}
 \caption{Second test measurement of purification with the electrostatic filter F1 in extraction mode checked with the detector D2. Upper panel: detector D2 current (blue), electrostatic filter F1 current (red). Lower panel: Electrostatic filter voltage (green), extraction rate (black). The voltage applied to detector D2 is 200\,V, instead of the 9\,V applied in the measurements before. This results in higher currents measured at the detector. The delay of the fast component of the detector current drop is larger compared to figure 7, because of a slower circulation speed.\label{fig:extraction}}
\end{figure}

The next step was to check whether we can only temporarily store impurities in the electrostatic filter or also remove them from the TMBi, as well as to learn something about the trapping mechanisms. For this purpose, a small flow \SP{($\cal{O}$(2)\,ml/min)} of TMBi was continuously extracted at the anode opening during operation of the electrostatic filter \SP{(extraction rate in figure \ref{fig:extraction})}. Our expectation was that we should  be able to remove part of the charged impurities contained in the liquid. Unfortunately, our test was doubly restricted. Firstly, we could only remove a small part of the liquid because we were required to work with a limited amount of TMBi in the apparatus due to safety concerns (see also~\cite{Gerke:2022otq}). A second drawback was that the geometry of the extraction port was designed primarily for the removal of impurities created by pick-up of  electrons injected at the photocathode, while we actually started with a homogeneous distribution of charged impurities.
\\Figure \ref{fig:extraction} shows a measurement where a small TMBi extraction flow was switched on for a limited time during the measurement. Even though the measurement parameters (higher detector voltage, lower circulation flux) were slightly different, the same features can be seen as in figure~\ref{fig: FirstFilter}: A fast drop when the electrostatic filter is switched on and a fast rise when the electrostatic filter is switched off again, delayed by the circulation time from the electrostatic filter to the detector. Switching on the TMBi extraction at the anode with a flow of 0.04 ml/s, on the other hand, shows no influence on either the detector current or the filter current.\\
The fact, that we do not see any sizeable effect by the extraction, was somewhat expected, because removal is only efficient for impurities that accumulate in front of the anode opening, as would be the case for ions created by photo-electrons from the cathode that are then drifted to the anode. 
For a large and homogeneous concentration of ionic impurities in the liquid, that has been shown to exist in our apparatus in section~\ref{sec:methods}, the amount of ions removed at the anode is negligible.
Furthermore, we do not expect that impurities bound electrostatically on the insulators in the filter can be removed by our extraction mode. 

\subsection{Electropurification test with simplified geometry}
\begin{figure}[b]
 \centering 
 \includegraphics[width=.95\textwidth]{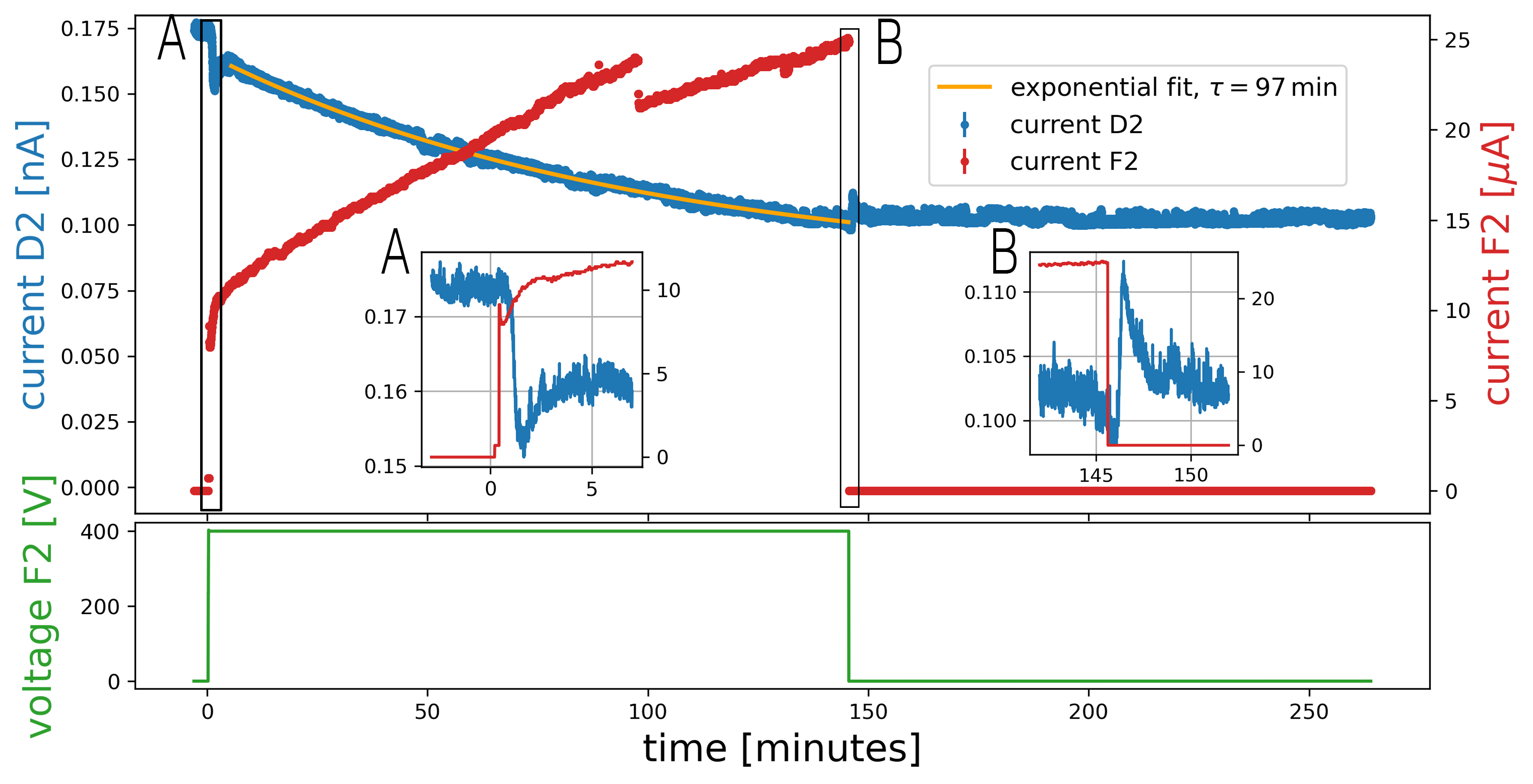}
 \caption{\label{fig:detector_as_filter} Test measurement of purification with the electrostatic filter F2, featuring a large ceramic surface opposite to a positively charged anode, checked with the detector D2. Upper panel: current measured at D2 and F2. lower panel: voltage applied to F2.}
\end{figure}

Since extracting larger amounts of TMBi is a problem in our application, as the next step we investigated whether the hypothesis of electrostatically retaining charged impurities on ceramic surfaces can be used as a purification method. 
For this purpose we use our first detector cell (D1), which contains a large ceramic surface around its anode, as electrostatic filter (F2).
By changing the mode of operation, grounding the cathode (shown red in figure~\ref{fig: detector scheme}) and the guard ring (blue) while applying a positive high voltage to the anode (green), an electric field similar to that of a coaxial cable is generated. TMBi is then pumped through the interfaces between the anode, the ceramic and the guard ring. Again, the current in the neighbouring detector D2 is recorded as a measure of purification.\\
Figure \ref{fig:detector_as_filter} shows the data of this investigation. 
After the initial circulation, the voltage at the filter F2 is ramped up from 0\,V to 400\,V at $t=0$ with a ramping speed of 50\,V/s. 
The same behaviour as before is observed after ramp up: Approximately 30 seconds after the maximum voltage  is reached the detector current drops within 30\,s from 175\,pA to 155\,pA. After a partial recovery of the current a much slower further decrease is observed with a time constant of about 80\,min down to about 110\,pA. Again, the filter current in F2 shows the opposite behaviour.\\
But one feature is very different to the previous measurements: The detector current remains low at about 110\,pA after the filter voltage has been switched off. 
It seems that charged impurities got attached on the ceramic between anode and guard tube in F2 by the electric field and were kept in place there even after switching off the high voltage. This retention appears to be long-lasting, at least on the observed time scale of about 120~minutes.\\
\SP{According to equation \ref{eq:conductivity} the reduction of the detector current by a factor 1.6 corresponds to a reduction of the impurity concentration by a factor of 1.6-2.6, depending on underlying process that generates the ions (equations \ref{eq:1nach2}-\ref{eq:1uLnach1uL})}.
In this third experiment we found for the first time a possibility to achieve a lasting cleaning effect in TMBi with the help of an electrostatic filter.
This possibility should be pursued further in the future.
By disconnecting the F2 filter from the apparatus and removing the impurities, it should be possible to develop this effect as a sustainable cleaning method. 
The geometry of the interfaces between ceramics and electrodes should also be further optimised in order to achieve the highest possible filter effect. 
\section{Permittivity measurement} \label{sec:relativePermittivity}
One of the essential quantities for the design of a TMBi based PET detector is the relative permittivity \epsr\ of the liquid which, together with the geometry, determines the capacitance of the device. It is also required to calculate the free electron yield following Onsagers theory~\cite{Onsager} which is an important quantity to characterize the performance of the detector. The relative permittivity has not been measured previously for TMBi.\\
In this section we present the method we chose for the determination of \epsr\ with our setup and the analysis of our measurements yielding the relative permittivity of liquid TMBi. 
\subsection{Method of relative permittivity determination}
\begin{figure}[htbp]
\centering 
\includegraphics[width=.95\textwidth]{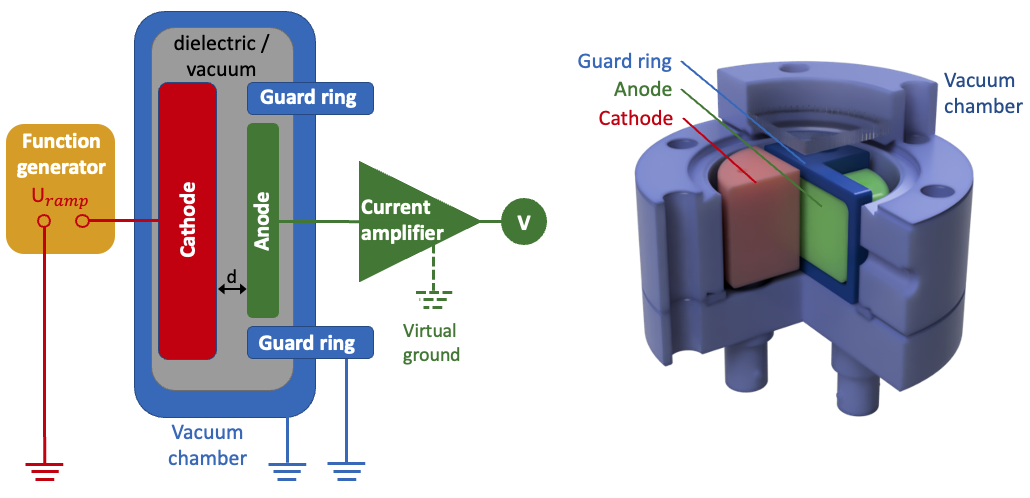}
\caption{\label{plt:measurment_setup_permittivity}Schematic drawing of the setup used for the determination of \epsr. Left: electric connection diagram. The voltage on the cathode is supplied by a Rigol DG4162 waveform generator. In distance $d$ to the cathode, the anode is being closely surrounded by a grounded guard ring and connected to virtual ground by a FEMTO-DDPCA300 current amplifier. It converts the current induced on the anode with a gain of $10^8$\,V/A to a voltage, which is then measured by a MSO 4054 digital oscilloscope. The volume between the electrodes can be evacuated or filled with a dielectric liquid. Right: CAD-drawing of detector D3, which is housed in a CF40 vacuum chamber.}
\end{figure}

To determine the permittivity, we compare capacitance measurements taken with an evacuated detector cell to measurements taken with a TMBi filled detector (D3). The ratio of the two capacitance values results in the permittivity of TMBi. This only holds, if the capacitance of the detector is measured and not influenced by stray capacities. \\
Figure \ref{plt:measurment_setup_permittivity} shows the setup developed for this purpose. 
In the measurement we apply a variable voltage to the cathode and measure the induced current on the anode using a high sensitivity FEMTO-DDPCA300 current amplifier. To minimize the influence of stray capacities in the setup the surface area of the anode is smaller than the cathode and surrounded by a guard ring on ground potential. The anode itself is connected to a virtual ground provided by the amplifier.
Compared to the geometry of the detector D2 presented in section \ref{sec: description_detector}  detector D3 features a significantly larger capacitance because of the larger dimensions of the anode as shown in Tab.~\ref{tab:detectors}.\\
We have successfully applied the above described method with a similar setup~\cite{master_simon_peters} to determine the relative permittivity of TMS (Tetramethyl-Silane), reproducing the value from literature with a relative deviation of less than 1\,\%. 
\begin{table}[h]
\centering
\caption{Comparison of the dedicated permittivity measurement setup and detector D2 presented in section \ref{sec: description_detector} \label{tab:detectors}.}
\begin{tabular}{|l|r|r|}
\hline
 & detector D2 & detector D3  \\ \hline
surface area {[}mm$^2${]} & 23 & 500 \\ \hline
distance anode to guard ring {[}mm{]} & 0.5 & 0.5 \\ \hline
plate distance {[}mm{]} & 1 & 2.1 \\ \hline
total volume {[}ml{]} & 12 & 23 \\ \hline
active volume {[}ml{]} & 0.023 & 1 \\ \hline
\end{tabular}
\end{table}
\subsection{Calculation of induced anode current}
The charge $Q(t)$ induced on the anode is connected to the capacitance of the setup and the detector voltage by 
\begin{equation}
    Q(t) = C \cdot U (t)
\end{equation}
Therefore a varying voltage $U(t)$ at the detector leads to a time-dependent current of: 
\begin{equation} \label{eq:i_c}
    I_C(t) = \frac{\partial Q(t)}{\partial t} = C \cdot \frac{\partial U(t)}{\partial t} = C \cdot \dot U
\end{equation}
To determine the capacitance $C$ we ramp the voltage $U(t)$ at the cathode of the detector  periodically up and down with a period $T$ and a constant ramping speed $\dot U_0$ as seen in the lower panel of figure \ref{plt: permit}. $U(t)$ can be described over one period by:
\begin{equation}
  U(t) =
    \begin{cases}
      +\dot U_0 \cdot \bigl( t-\frac{T}{4}\bigr) & \text{for $0 \leq t < \frac{T}{2}  $}\\
      - \dot U_0 \cdot \bigl( t-\frac{3 \cdot T}{4}\bigr) & \text{for $\frac{T}{2} \leq t < T $}\\
    \end{cases}       
\end{equation}
According to Eq. (\ref{eq:i_c}) this ramping voltage corresponds to a periodic current with a rectangular shape in the ideal case:
\begin{equation}
  I_C(t) =
    \begin{cases}
      +C\cdot \dot U_0 & \text{for $0 \leq t < \frac{T}{2} $}\\
      - C \cdot \dot U_0 & \text{for $\frac{T}{2}  \leq t < T $}\\
    \end{cases}       
\end{equation}
For the detector filled with TMBi we expect in addition to the current $I_C(t)$, describing the charging/discharging of the detector capacitance $C$, a leakage current $I_\mathrm{leakage}(t)=U(t)/R$ due to the finite resistance $R$ of the TMBi yielding to a total current of 
\begin{equation}
 I_\mathrm{tot}(t) = I_C(t)+I_\mathrm{leakage}(t) = 
    \begin{cases}
      + \dot U_0 \cdot \left( C + \frac{1}{R}\cdot \bigl( t-\frac{T}{4}\bigr) \right) & \text{for $0 \leq t < \frac{T}{2} $}\\
      - \dot U_0 \cdot \left( C + \frac{1}{R}\cdot \bigl( t-\frac{3 \cdot T}{4}\bigr) \right) & \text{for $\frac{T}{2}  \leq t < T $}\\
    \end{cases}       
\end{equation}
The total current $I_\mathrm{tot}(t)$ then corresponds to the sum of a periodic rectangular function and a sawtooth function which can be expressed by a real or a complex Fourier series (using only odd numbered coefficients for these two special periodic functions):
\begin{eqnarray} 
 I_\mathrm{tot}(t) &=& \sum_{k=0}^\infty a_{2k +1} \cdot \cos \biggl( (2k +1) \cdot \omega_0 \cdot t\biggr) + b_{2k +1} \cdot \sin \biggl( (2k +1) \cdot \omega_0 \cdot t\biggr) \nonumber \\
 &=& \sum_{k=-\infty}^\infty A_{2k +1} \cdot \exp{\biggl(i \cdot (2k+1) \cdot \omega_0 \cdot t \biggr)} \label{eq:Fourier_series}
\end{eqnarray}
\begin{eqnarray}
    \mathrm{with}\quad \omega_0 = \frac{2\pi}{T} \quad &\mathrm{and}& \quad a_{2k +1} = -\frac{\dot U_0 \cdot T}{R} \cdot \frac{2}{\pi^2 \cdot (2k+1)^2} \qquad \mathrm{and} \quad 
    b_{2k +1} = \dot U_0 \cdot C \cdot \frac{4}{\pi \cdot (2k+1)}
    \nonumber \\ \qquad &\mathrm{and}& \quad 
     A_{2k +1} = \frac{ a_{2k +1} - i \cdot  b_{2k +1}}{2} = \bar A_{-(2k+1)}
\end{eqnarray}
The detector current is measured with a FEMTO-DDPCA300 current amplifier which possesses at the input a low pass filter with the time constant $\tau$, corresponding to an edge frequency $f_\mathrm{c} = \frac{1}{2\pi\tau}$, and a corresponding transfer function in Fourier space $G(\omega)=\frac{1}{1+i\cdot \omega \cdot \tau}$, which also accommodates a finite time constant of our voltage supply.
The measured current $I^\mathrm{meas}_\mathrm{tot}(t)$ is then the product of the complex Fourier series from Eq. (\ref{eq:Fourier_series}) and the transfer function $G(\omega)$ yielding in the notation of a real Fourier series:
\begin{equation}
     I^\mathrm{meas}_\mathrm{tot}(t) = \sum_{k=0}^\infty a^\prime_{2k +1} \cdot \cos \biggl( (2k +1) \cdot \omega_0 \cdot t\biggr) + b^\prime_{2k +1} \cdot \sin \biggl( (2k +1) \cdot \omega_0 \cdot t\biggr) \label{eq:fit_function}
\end{equation}
\begin{eqnarray}
    &\mathrm{with}& \quad a^\prime_{2k +1} = 
    \biggl( a_{2k +1} - b_{2k +1} \cdot (2k+1)  \cdot \omega_0 \cdot \tau \biggr) \cdot 
    \frac{1}{1+\bigl( (2k+1) \cdot \omega_0 \cdot \tau \bigr)^2} \nonumber \\
   &\mathrm{and}& \quad b^\prime_{2k +1} = 
    \biggl( b_{2k +1} + a_{2k +1} \cdot (2k+1)  \cdot \omega_0 \cdot \tau \biggr) \cdot 
    \frac{1}{1+\bigl( (2k+1) \cdot \omega_0 \cdot \tau \bigr)^2}
\end{eqnarray}
\subsection{Measurement of relative permittivity of liquid TMBi}
\begin{figure}[t]
\centering 
\includegraphics[width=.95\textwidth]{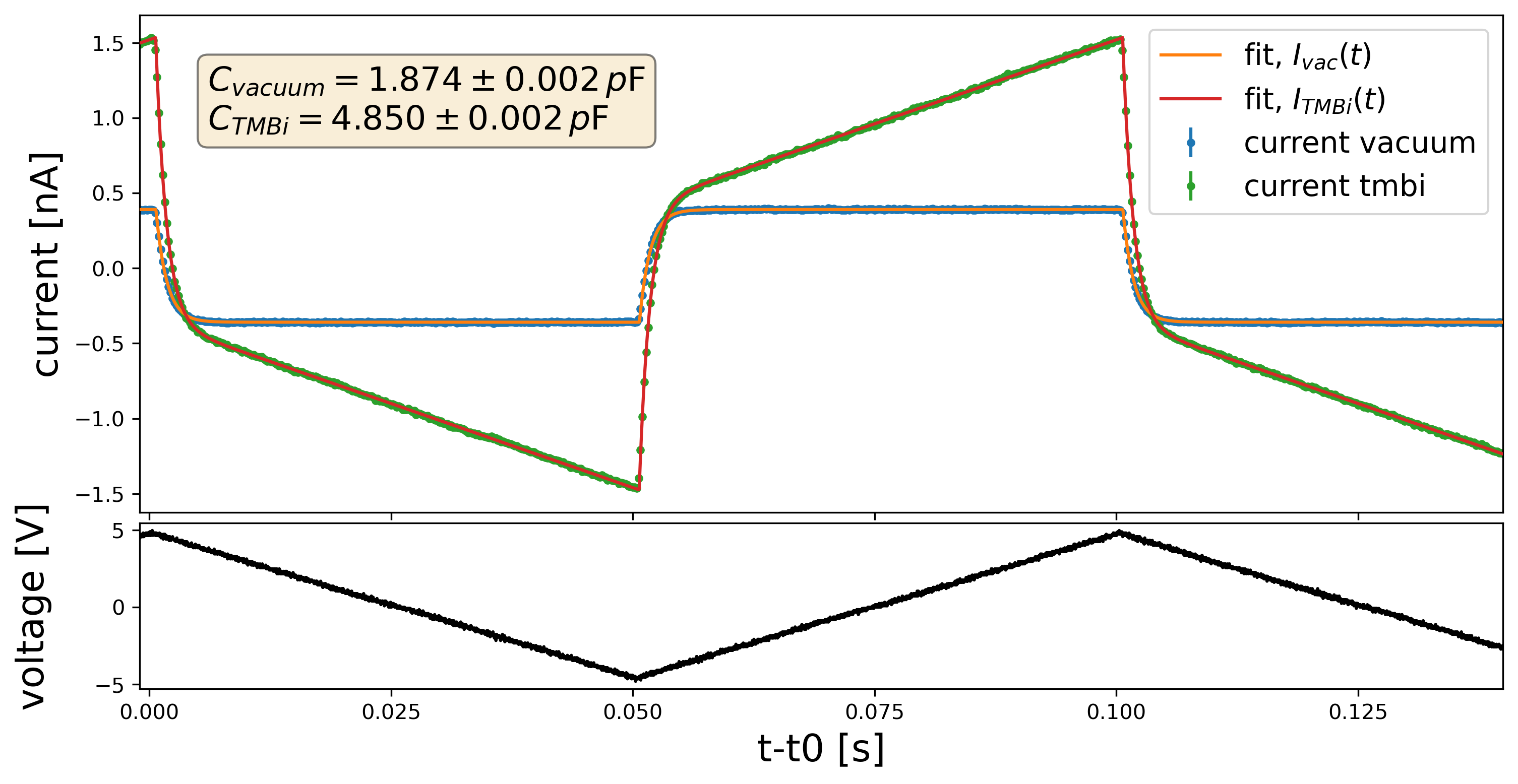}
\caption{\label{plt: permit}Permittivity measurement.  Upper panel: current measured in the detector, filled with TMBi (green) and under vacuum (blue) and the fit applied to both data sets (red). Lower panel: voltage applied to the cathode.}
\end{figure}

Using detector D3, the current response of the anode to a voltage ramp applied to the cathode has been measured. The period of this triangular ramp was $T=100\,\mathrm{ms}$ and the ramping speed was $\dot U_0=\pm 200\,\mathrm{V/s}$ corresponding to an amplitude of $\pm 5\,\mathrm{V}$.
Figure~\ref{plt: permit} shows the currents measured 
first in the evacuated setup and then 
with the measurement cell filled with TMBi,
together with the fitted current responses according to Eq.~(\ref{eq:fit_function}). 
The ratio of the  capacities of the measurement cell filled with TMBi and under vacuum obtained from the two fits results in the relative permittivity of TMBi:
\begin{equation}
    \epsilon_r = \frac{C_{\rm TMBi}}{C_{\rm vacuum}} = 2.59 \pm 0.003_{\rm stat} \pm 0.026_{\rm sys}
\end{equation}
Here, the statistical uncertainty is provided by the fit, while the systematic uncertainty is estimated from our earlier TMS measurements to be on the order of 1\% of the measured value~\cite{master_simon_peters}. The measured value is found to be consistent with the assumptions made based on the refractive index of TMBi during previous free ion yield measurements on TMBi \cite{Farradeche}.

\section{Conclusion}
Metal-organic compounds have long been used for ionisation detectors. TMBi, which contains the heavy element bismuth (Z=83), has been proposed as a detector medium with Cherenkov light and ionisation readout for the measurement of 511 keV annihilation quanta in a novel detector for position emission tomography. One difficulty in this detector design is the safe handling of TMBi~\cite{Gerke:2022otq}, another challenge is the purification of TMBi from electro-negative impurities so that the resulting free electrons in TMBi can be drifted to the anode without losses.\\
In this article, we describe the design of a dedicated test setup with a electrostatic filter for cleaning TMBi that offers free electrons \SP{from a photocathode} to electronegative impurities in TMBi, converting them into negative ions and drifting them electrically to an anode where they eventually can be removed. For this setup we designed and built a small volume miniaturized magnetically driven piston pump that is also presented in this paper. We investigated the purification success by measuring the change in dark current in an adjacent custom built ionisation detector through which the TMBi was circulated by a pump. \\
\SP{The proposed electrostatic filter mechanism requires the injection of free electron densities in the order of 1\,$\mu$A/cm$^2$ to be reasonably fast. Of course if a precleaning of the TMBi can be done with other purification methods or 
the reappearance of impurities, e.g. by outgassing, is slow enough, smaller current densities are possible.}
We started with contaminated TMBi, which unfortunately already contained so many \SP{impurities, which dissociated into ions under high voltage,} that we refrained from injecting free electrons with the photocathode built into the electrostatic filter. \\
Our first experiments showed that this method in principle works. We achieved a reduction of the detector current and thus a reduction of the impurities by applying drift fields at the level of 4~kV/cm. The two different time constants of the reduction indicated two different cleaning mechanisms. While the one with a time constant of about 15 seconds pointed to the intended cleaning by drifting the charged impurities to the anode, we interpreted the significantly longer time constants in the ${\cal O}(100)$\,min as electrostatic accumulation of charged impurities on ceramic surfaces. Both were further investigated in dedicated measurements. Unfortunately, the accumulation at the aznode could not be investigated specifically with the apparatus, because the starting impurity was so large that the locally targeted injection of photo-electrons had too little influence compared to the charge carriers already present in the TMBi, and therefore only a small proportion of the impurities were drifted to the planned extraction point. In contrast, the hypothesis of accumulation of ionic impurities at the ceramics could be further strengthened with a measurement using a dedicated electrostatic precipitator. The latter cleaning effect persisted after the high voltage was switched off. In future measurements, we want to try to use this effect in a targeted manner for multi-stage cleaning. One possibility would be to separate the volume with the ceramic and impurities deposited on it, purge the contained material to obtain a clean state again and then to use it for a next cleaning step. \\
In the second part of the article we present the first direct measurement of the relative permittivity of TMBi to our knowledge. In one of our ionisation detectors, we were able to measure the current induced on the anode by a triangular voltage ramp on the cathode without any effect on stray capacities. We present the associated mathematical model in detail. By taking the ratio of the measured capacities with the detector filled with TMBi and under vacuum, we were able to determine the relative permittivity to be $\epsilon_r = 2.59 \pm 0.003_{\rm stat} \pm 0.026_{\rm sys}$.

\acknowledgments
This work is supported by Deutsche Forschungsgemeinschaft (DFG), project numbers WE 1843/8-1 and SCHA 1447/3-1, and French national research agency (ANR), project number ANR-18-CE92-0012-01
%


\begin{thebibliography}{99}
%
 \bibitem{Yvon2014} D. Yvon et al.,
 \emph{CaLIPSO: An Novel Detector Concept for PET Imaging}, 
 \href{https://doi.org/10.1109/TNS.2013.2291971}{IEEE Transactions on Nuclear Science 61 (2014) 60-66}
%
\bibitem{ramos} E. Ramos et al.,
 \emph{Trimethyl Bismuth Optical Properties for Particle Detection and the CaLIPSO Detector},
 \href{https://doi.org/10.1109/TNS.2015.2424080}{IEEE Transactions on Nuclear Science 62 (2015) 1326-1335}
%
\bibitem{Canot} C. Canot et al.,
 \emph{Fast and efficient detection of 511~keV photons using Cherenkov light in PbF$_2$ crystal, coupled to a MCP-PMT and SAMPIC digitization module}, 
 \href{https://dx.doi.org/10.1088/1748-0221/14/12/P12001}{JINST 14 (1019) P12001-P12001} 
%
\bibitem{calipso_sym} O. Kochebina et al., 
 \emph{Performance Estimation for the High Resolution CaLIPSO Brain PET scanner: A Simulation Study},
 \href{https://doi.org/10.1109/TRPMS.2018.2880811}{IEEE Trans. Radiat. Plasma Med. Sci. 3 (2019) 363–370} 
%
\bibitem{xcom}
S. Seltzer, \emph{Xcom-photon cross sections database}, \href{https://physics.nist.gov/PhysRefData/Xcom/html/xcom1.html}{nist standard reference database 8,1987. Accessed: 2022-09-09.}
%
\bibitem{Holroyd}
Richard A. Holroyd et al.,
\emph{Electron attachment to oxygen and other solutes in non-polar liquids},
\href{https://doi.org/10.1016/0146-5724(80)90143-0}{Radiation Physics and Chemistry (1977),
Volume 15, Issues 2–3,1980.}
%
\bibitem{Martens} Kai Martens, Xenon Dark Matter Project, private communication 
%
\bibitem{Gerke:2022otq} B.~Gerke et al.,
 \emph{Suppression of electrical breakdown phenomena in liquid TriMethyl Bismuth based ionization detectors},
 \href{https://doi.org/10.48550/arXiv.2206.13440}{arXiv:2206.13440 [physics.ins-det]}
\bibitem{Brown} Brown et al.,
 \emph{Magnetically-coupled piston pump for high-purity gas applications},
 \href{https://doi.org/10.1140/epjc/s10052-018-6062-z}{The European Physical Journal C 78 (2018) 604}
%
\bibitem{Farradeche}
Farrad{\`e}che et al.,
 \emph{Free ion yield of trimethyl bismuth used as sensitive medium for high-energy photon detection},Journal of Instrumentation, 13 P11004, 2018
%
\bibitem{master_simon_peters} S. Peters,
 \emph{Set Up of a Purification and Measurement System for Organometallic Liquids for a Future PET Detector and Test Measurements with Tetramethylsilane},
 Masters Thesis, University of Münster, 2019
%
\bibitem{Staudhammer}
P. Staudhammer et al.,
\emph{Electrical resistivity of cyclohexane as a function of
temperature and water concentration},
\href{https://doi.org/10.1063/1.1722763}{Journal of Applied Physics, vol. 28, pp. 405–410, Apr. 1957.}
%
\bibitem{liquid-state-properties}
Kunhardt, E E and Christophorou, L G and Luessen, L H
\emph{The liquid state and its electrical properties},
\href{https://link.springer.com/chapter/10.1007/978-1-4684-8023-8_20}{Springer, (1989), NATO Science Series: B}
%
\bibitem{Teflon} Abhijit Nag et al.,
 \emph{Tribochemical Degradation of Polytetrafluoroethylene in Water and Generation of Nanoplastics},
 \href{https://doi.org/10.1021/acssuschemeng.9b03573}{ACS Sustainable Chem. Eng. 7 (2019) 17554-17558}
%
\bibitem{schmidt} Werner F. Schmidt,
 \emph{Liquid State Electronics of Insulating Liquids},
 CRC Press (1997) ISBN 13: 9780849344459
%
\bibitem{patent} H. Niedermeyer, M. Daelman et al.,
 \emph{Herstellung von trialkylpnictogen},
 \href{https://worldwide.espacenet.com/patent/search/family/062791590/publication/EP3587430A1?q=EP3587430A1}{Lanxess organimetallics GmbH., published as: EP3587430A1;EP3587430B1}
%
\bibitem{Onsager} L. Onsager,
 \emph{Initial Recombination of Ions},
 \href{https://doi.org/10.1103/PhysRev.54.554}{Phys. Rev. 54 (1938) 554-557}





\end{thebibliography}
\end{document}